\documentclass[floatfix,twocolumn,showpacs,prl,amsmath]{revtex4}
\usepackage{graphicx}
\usepackage{bm}
\usepackage{amssymb}
\usepackage{dcolumn}
\usepackage{subfigure}
\usepackage{threeparttable}
\usepackage{multirow}
\usepackage[usenames,dvipsnames]{color}
\usepackage{mathtools}

\begin{document}
\newcommand{\vn}[1]{{\boldsymbol{#1}}}
\newcommand{\polarivec}{\boldsymbol{\epsilon}}
\newcommand{\rmd}{{\rm d}}
\newcommand{\rme}{{\rm e}}
\newcommand{\vht}[1]{{\boldsymbol{#1}}}
\newcommand{\matn}[1]{{\bf{#1}}}
\newcommand{\matnht}[1]{{\boldsymbol{#1}}}
\newcommand{\bege}{\begin{equation}}
\newcommand{\ee}{\end{equation}}
\newcommand{\bal}{\begin{aligned}}
\newcommand{\defbar}{\overline}
\newcommand{\SM}{\scriptstyle}
\newcommand{\eal}{\end{aligned}}
\newcommand{\torkance}{\vartheta}
\newcommand{\udot}{\overset{.}{u}}
\newcommand{\exponential}[1]{{\exp(#1)}}
\newcommand{\phandot}[1]{\overset{\phantom{.}}{#1}}
\newcommand{\phandag}{\phantom{\dagger}}
\newcommand{\Trace}{\text{Tr}}
\newcommand{\Bxc}{\Omega}
\setcounter{secnumdepth}{2}
\title{Laser-induced torques in metallic ferromagnets}
\author{Frank Freimuth}
\email[Corresp.~author:~]{f.freimuth@fz-juelich.de}
\author{Stefan Bl\"ugel}
\author{Yuriy Mokrousov}
\affiliation{Peter Gr\"unberg Institut and Institute for Advanced Simulation,
Forschungszentrum J\"ulich and JARA, 52425 J\"ulich, Germany}
\date{\today}
\begin{abstract}
We study laser-induced torques in 
bcc Fe, hcp Co and $L1_0$ FePt based on first-principles electronic structure calculations
and the Keldysh nonequilibrium formalism.
We find that the torques have two contributions, one from the
inverse Faraday effect (IFE) and one from the
optical spin-transfer torque (OSTT).  Depending on the ferromagnet
at hand and on the quasiparticle broadening the two contributions may
be
of similar magnitude or one contribution may dominate over the other.
Additionally, we determine the nonequilibrium spin polarization in order
to investigate its relation to the torque.
We find the torques and the perpendicular component of the 
nonequilibrium spin polarization to be odd in the helicity of the
laser light, while the spin polarization that is induced parallel to
the
magnetization is helicity-independent. The parallel component of the
nonequilibrium spin polarization is orders of magnitude larger than the
perpendicular component.
In the case of hcp Co we find
good agreement between the calculated laser-induced torque and a
recent experiment.
\end{abstract}


\maketitle
\section{Introduction}
Several mechanisms induce torques on the
magnetization in magnetically ordered materials when 
laser pulses are applied~\cite{rmp_ultrafast}.
When circularly polarized light is used, an effective magnetic
field parallel to the light wave vector acts on the magnetization
due to the inverse Faraday effect 
(IFE)~\cite{Kimel_ultrafast_control_magnetization} (Figure 1a).
The IFE is thought to play a crucial role in the
laser-induced magnetization reversal in ferromagnetic thin 
films~\cite{Lambert_all_optical_control,magnetization_switching_FePt}.
Additionally, there is a light-induced effective magnetic field 
perpendicular to both the magnetization
and the light wave vector, which leads to the 
optical spin transfer torque (OSTT)~\cite{nemec_ostt} (Figure 1b). 
Besides these non-thermal effects, the laser-induced heating can
also generate torques due to heat-induced modifications of the
magnetic anisotropy 
fields~\cite{Kampen_optical_probe_coherent_spin_waves}.
Furthermore, laser pulses excite superdiffusive
spin currents in magnetic
heterostructures~\cite{battiato_superdiffusive_spin_transport_ultrafast_demagnetization,
Malinowski_ultrafast_demag_direct_spin_transfer,
melnikov_AuFeMgO_PhysRevLett.107.076601,
thz_emitter_Seifert}, which 
mediate spin-transfer torques when
they flow from one magnetic layer into 
another~\cite{ultrafast_stt_laser}.
Finally, the laser-induced heating drives spin-currents due to the
spin-dependent Seebeck effect, which leads to thermal spin transfer
torques in metallic spin-valves~\cite{choi_thermal_stt}.

\begin{figure}
\includegraphics[width=\linewidth,trim=11cm 20.5cm 1cm 2cm,clip]{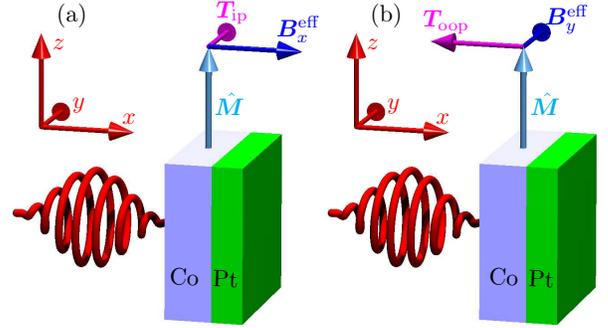}
\caption{\label{figure1}
A circularly polarized light pulse propagates in $x$ direction and hits a 
Co/Pt bilayer. The magnetization direction $\hat{\vn{M}}$ is 
along the $z$ axis. The laser-induced torque $\vn{T}$ has two components:
(a) The in-plane component $\vn{T}_{\rm ip}$ can be attributed to an effective
magnetic field $\vn{B}^{\rm eff}_{x}$ in $x$ direction. $\vn{T}_{\rm ip}$ leads to an initial in-plane 
tilt of $\hat{\vn{M}}$.
(b) The out-of-plane component $\vn{T}_{\rm oop}$ can be attributed to the
$y$ component $\vn{B}^{\rm eff}_{y}$ of a laser-induced 
effective magnetic field and leads to an initial out-of-plane tilt
of $\hat{\vn{M}}$.
}
\end{figure}

In the following we will consider only the effective magnetic fields,
torques and non-equilibrium spin-densities related to the IFE and OSTT.
In ferromagnets the light-induced non-equilibrium spin-density
can generally exhibit a component parallel to the equilibrium
magnetization as well as a perpendicular one. The perpendicular
component exerts a torque on the magnetization and tilts it.
This laser-induced torque has been investigated in metallic ferromagnets
in recent 
experiments~\cite{choi_thesis,femtosecond_control_electric_currents_Huisman}:
In Co a 50~fs laser pulse with fluence 1~mJ~cm$^{-2}$ induces an
effective magnetic field whose perpendicular component 
has been estimated at 
0.2 Tesla. One experiment~\cite{choi_thesis} was interpreted
in terms of an initial out-of-plane tilting of the
magnetization due to an
out-of-plane torque (Figure 1b), while the 
second experiment~\cite{femtosecond_control_electric_currents_Huisman} 
was interpreted in terms of an 
initial in-plane tilting due to an
in-plane torque (see Figure 1a).
The out-of-plane tilting has been ascribed to the OSTT
and an in-plane tilting is expected from the
IFE. 
Both experiments find that the magnetization is only tilted when 
circularly polarized light is used and that the effect changes sign
when the helicity of the light is reversed.
In both experiments the Co layer is sufficiently thick (10nm) to assume that
the laser-induced effective magnetic fields responsible for 
the magnetization tilting can be modelled theoretically based on the bulk
electronic structure of Co, neglecting the Co/Pt interface.
In one experiment~\cite{choi_thesis} the Pt capping layer 
mainly serves to prevent oxidation of the Co layer.
In the second
experiment~\cite{femtosecond_control_electric_currents_Huisman} 
the inverse spin-orbit torque (ISOT)~\cite{invsot} due to the
structural
inversion asymmetry at the Co/Pt interface 
is exploited to convert the magnetization tilting into an interfacial
photocurrent. 

On the theory side, 
for the special case of light-propagation direction parallel to the
magnetization, light-induced effective magnetic fields parallel
to the magnetization have been studied in 
transition metal ferromagnets~\cite{ife_berritta} 
with \textit{ab-initio} methods
as well as in  
the ferromagnetic Rashba model~\cite{light_induced_magnetic_field_rashba}.  
Both theoretical works find that not only circularly polarized light
but also linearly polarized light induces 
effective magnetic fields parallel to the magnetization. Moreover, it
was found that the light-induced spin polarization parallel to the
magnetization is almost helicity-independent in Fe, Co, and Ni~\cite{ife_berritta}. 
Since in contrast the light-induced torques observed experimentally are odd in the
helicity~\cite{femtosecond_control_electric_currents_Huisman} is seems
that effective magnetic 
fields perpendicular to the
magnetization direction depend 
differently on the light helicity than
the parallel component in these metallic ferromagnets.

In this work we use \textit{ab-initio} density
functional theory in order to
study all components of
the light-induced non-equilibrium spin density and of the 
resulting torques and effective magnetic fields in Fe, Co and FePt.
This allows us to answer the two questions raised above: 
(i) Is the laser-induced torque on the magnetization
in Figure 1 pointing in the in-plane or in the
out-of-plane direction? (ii) How do the parallel and perpendicular
components of the light-induced effective magnetic field
differ regarding their size and their dependence on the light
polarization?

This paper is structured as follows:
In section~\ref{sec_computational_method} we describe our 
computational approach, which uses the Keldysh non-equilibrium
formalism to obtain the response in second order to the electric field
of the laser. Details of the derivation and of the numerical
implementation are given in appendix~\ref{sec_formalism}
and in appendix~\ref{sec_implementation}, respectively.
Before presenting our results 
in section~\ref{sec_results} we first describe the computational
parameters used in the calculations in 
\ref{sec_computational_parameters}.   
In section~\ref{sec_laser_induced_torques}
we discuss the effective magnetic fields that give rise to
the laser-induced torques and in section~\ref{sec_laser_induced_spin}
we investigate the laser-induced nonequilibrium spin density.  
We conclude with a summary in section~\ref{sec_summary}.
\section{Computational method}
\label{sec_computational_method}
We use Kohn-Sham density functional theory to describe 
interacting many-electron systems by the effective single-particle
Hamiltonian
\bege\label{eq_ks_hamil}
H(\vn{r})=H_{0}(\vn{r})-\vn{m}\cdot\hat{\vn{M}}\Bxc^{\rm xc}(\vn{r}),
\ee
where $H_0$ contains kinetic energy, scalar potential and 
spin-orbit interaction (SOI), $\vn{m}=-\mu_{\rm B}\vht{\sigma}$
is the spin magnetic moment operator, $\mu_{\rm B}$ is the Bohr magneton, 
$\vht{\sigma}=(\sigma_x,\sigma_y,\sigma_z)^{\rm T}$ is the vector of Pauli spin matrices,
$\hat{\vn{M}}$ is a normalized vector parallel to the magnetization,
$\Bxc^{\rm xc}(\vn{r})=\frac{1}{2\mu_{\rm B}}[V^{\rm eff}_{\rm
  minority}(\vn{r})-V^{\rm eff}_{\rm majority}(\vn{r}) ]$ is 
the exchange 
field, and $V^{\rm eff}_{\rm minority}(\vn{r})$ and $V^{\rm eff}_{\rm majority}(\vn{r})$ are the
effective potentials of minority and majority electrons, respectively.

The interaction with the laser field is described by the perturbation to the Hamiltonian 
\bege\label{eq_laser_perturbation}
\delta H(t)=e\vn{v}\cdot\vn{A}(t),
\ee
where $e$ is the elementary positive charge, $\vn{v}$ is the velocity operator
and 
\bege
\vn{A}(t)=
{\rm Re}
\left[
\frac{E_{0}\polarivec e^{-i\omega t}}{i\omega}
\right]
\ee
is the vector potential.
The corresponding electric field of the laser is
\bege
\vn{E}(t)=-\frac{\partial \vn{A}(t)}{\partial t}={\rm Re}
\left[
E_{0}\polarivec e^{-i\omega t}
\right],
\ee
where $\polarivec$ is the light-polarization vector and $E_0$ is the amplitude of the
electric field. We assume that $E_0$ is real-valued. However, $\polarivec$ may be complex.
For example, to describe left-circularly and right-circularly polarized light propagating in $x$ direction
we use $\polarivec=(0,1,i)/\sqrt{2}$ and $\polarivec=(0,1,-i)/\sqrt{2}$, respectively.

The laser-induced change of spin polarization is 
given
by~\cite{light_induced_magnetic_field_rashba,spin_polarization_ti_murakami,
tatara_PhysRevB.84.174433,tatara_PhysRevLett.109.127204}
\bege\label{eq_define_delta_s}
\delta\vn{S}=\int \rmd^3 r \,\delta\vn{s}(\vn{r})=\frac{\hbar}{2i}{\rm Tr}
\left[
\vn{\sigma}G^{<}
\right],
\ee
where
$G^{<}$ is the lesser Green function. $\delta\vn{S}$ is the 
integral of the nonequilibrium spin-density $\delta \vn{s}(\vn{r})$ over the 
simulation volume, i.e., the change of
the total electron spin in the simulation volume, when $\hat{\vn{M}}$ in 
Eq.~\eqref{eq_ks_hamil}
is kept fixed. 
The torque on the magnetization due to the nonequilibrium spin-density is given
by~\cite{stt-cocuco_haney&waldron&duine&nunez&guo&macdonald,CurrentinducedtorquesinmagneticmetalsBeyondspintransfer_Haney_2008,invsot,ibcsoit}
\bege\label{eq_torque_from_spin_density}
\vn{T}=
\frac{2\mu_{\rm B}}{\hbar}
\int \rmd^3 r \,
\Bxc^{\rm xc}(\vn{r})
\delta\vn{s}(\vn{r})\times\hat{\vn{M}}.
\ee
Since the nonequilibrium
spin-density $\delta\vn{s}(\vn{r})$ as well as the exchange field $\Bxc^{\rm xc}(\vn{r})$ 
vary strongly on the atomic scale, 
it is generally not possible to calculate $\vn{T}$ exactly from $\delta\vn{S}$.
Therefore, we calculate the torque from 
\bege\label{eq_torque_from_lesser}
\vn{T}=i{\rm Tr}
\left[
\vn{\mathcal{T}}G^{<}
\right],
\ee
where $\vn{\mathcal{T}}(\vn{r})=\vn{m}\times\hat{\vn{M}}\Bxc^{\rm
  xc}(\vn{r})$ is the 
torque
operator~\cite{invsot,ibcsoit,CoPt_Haney_Stiles,torque_operator_turek,sot_ebert}.
It is clear that the laser-induced nonequilibrium magnetization in
paramagnets and diamagnets consists of both spin and orbital
contributions.
Consequently,
a recent \textit{ab-initio} study on the IFE considered both spin and orbital parts of
the laser-induced nonequilibrium magnetization~\cite{ife_berritta}. However, in the present work we are
mostly interested in the laser-induced torques on the magnetization in
ferromagnets, which are determined by
the nonequilibrium spin-density according to Eq.~\eqref{eq_torque_from_spin_density}. 
While the laser-induced
orbital polarization corresponds to orbital currents, which lead to magnetic fields according to the
Maxwell equations, the resulting torques are negligible in 
comparison to the torques described by
Eq.~\eqref{eq_torque_from_spin_density}. We 
therefore do not consider the laser-induced orbital polarization
in this work.

In systems with broken inversion symmetry, $\vn{T}$ contains a
contribution that is first order in $\vn{E}(t)$, the so-called spin-orbit
torque (SOT)~\cite{invsot,ibcsoit,CoPt_Haney_Stiles,sot_ebert,
symmetry_spin_orbit_torques,Ciccarelli_NiMnSb}. 
However, this first-order contribution oscillates with frequency $\omega$.
Since the light frequency $\omega$ is much higher than the ferromagnetic resonance 
frequency, this oscillating contribution will not induce
significant magnetization dynamics. Therefore, we consider
the dc part in the response to a continuous laser field. The contribution to $\vn{T}$ that is
second order in $\vn{E}(t)$ contains such static terms. They can arise for example from the
time-independent part $E_0^2 {\rm Re}[\epsilon_i \epsilon_j^*]/2$
in 
\bege
E_{i}(t)E_{j}(t)\!=\!\frac{E_0^2}{4}
\!\left[
\epsilon_i\epsilon^*_j
\!+\!
\epsilon^*_i\epsilon_j
\!+\!
\epsilon_i\epsilon_j e^{-2i\omega t}
\!+\!
\epsilon^*_i\epsilon^*_j e^{2i\omega t}
\right].
\ee
The dc correction of $G^{<}$ proportional to $E_0^2$ can be
conveniently derived within the Keldysh nonequilibrium formalism. 
Details of the derivation are given in Appendix~\ref{sec_formalism}.
The resulting torque is given by the expression
\bege\label{eq_define_torque_chi}
T_{i}=\frac{a_{0}^3 I}{c}
\left(
\frac{\mathcal{E}_{\rm H}}{\hbar\omega}
\right)^2
{\rm Im}
\sum_{jk}
\epsilon_j
\epsilon_k^*
\chi_{ijk},
\ee
where $c$ is the velocity of light,
$a_{0}=4\pi\epsilon_0\hbar^2/(me^2)$ is Bohr's radius,
$I=\epsilon_0 c E_{0}^2/2$ is the intensity of light,
$\epsilon_0$ is the vacuum permittivity and
$\mathcal{E}_{\rm H}=e^2/(4\pi\epsilon_0 a_0)$ is the Hartree energy.
The tensor $\chi_{ijk}$ is given by
\bege\label{eq_chi_noabrev}
\begin{aligned}
\chi_{ijk}=&\frac{2}{\mathcal{N}\hbar a_0^2 \mathcal{E}_{\rm H}}\sum_{\vn{k}}
\int \rmd \mathcal{E}
{\rm Tr}\Big[\\
&f(\mathcal{E})
\mathcal{T}_{i}
G^{\rm R}_{\vn{k}}(\mathcal{E})
v_{j}
G^{\rm R}_{\vn{k}}(\mathcal{E}-\hbar\omega)
v_{k}
G^{\rm R}_{\vn{k}}(\mathcal{E})
\\
-&
f(\mathcal{E})
\mathcal{T}_{i}
G^R_{\vn{k}}(\mathcal{E})
v_{j}
G^R_{\vn{k}}(\mathcal{E}-\hbar\omega)
v_{k}
G^A_{\vn{k}}(\mathcal{E})\\
+&f(\mathcal{E})
\mathcal{T}_{i}
G^{\rm R}_{\vn{k}}(\mathcal{E})
v_{k}
G^{\rm R}_{\vn{k}}(\mathcal{E}+\hbar\omega)
v_{j}
G^{\rm R}_{\vn{k}}(\mathcal{E})
\\
-&f(\mathcal{E})
\mathcal{T}_{i}
G^{\rm R}_{\vn{k}}(\mathcal{E})
v_{k}
G^{\rm R}_{\vn{k}}(\mathcal{E}+\hbar\omega)
v_{j}
G^{\rm A}_{\vn{k}}(\mathcal{E})
\\
+&
f(\mathcal{E}-\hbar\omega)
\mathcal{T}_{i}
G^R_{\vn{k}}(\mathcal{E})
v_{j}
G^R_{\vn{k}}(\mathcal{E}-\hbar\omega)
v_{k}
G^A_{\vn{k}}(\mathcal{E})\\
+&
f(\mathcal{E}+\hbar\omega)
\mathcal{T}_{i}
G^{\rm R}_{\vn{k}}(\mathcal{E})
v_{k}
G^{\rm R}_{\vn{k}}(\mathcal{E}+\hbar\omega)
v_{j}
G^{\rm A}_{\vn{k}}(\mathcal{E})
\Big],\\
\end{aligned}
\ee
where $\mathcal{N}$ is the number of $\vn{k}$ points used to
sample the Brillouin zone, $f(\mathcal{E})$ is the Fermi distribution function,
$G^{\rm R}_{\vn{k}}(\mathcal{E})$ is the retarded Green function and
$G^{\rm A}_{\vn{k}}(\mathcal{E})=[G^{\rm R}_{\vn{k}}(\mathcal{E})]^{\dagger}$ is the advanced Green function.
In order to simulate disorder and finite lifetimes of the electronic states
we use the constant broadening $\Gamma$. Therefore, the energy dependence of the Green function
is known analytically:
\bege \label{eq_define_green_analy}
G^{\rm R}_{\vn{k}}(\mathcal{E})=
\hbar
\sum_{n}\frac{|\vn{k}n\rangle\langle\vn{k}n|}{\mathcal{E}-\mathcal{E}_{\vn{k}n}+i\Gamma},
\ee
where $|\vn{k}n\rangle$ and $\mathcal{E}_{\vn{k}n}$ are eigenstates and eigenenergies, respectively, of
the Hamiltonian in Eq.~\eqref{eq_ks_hamil}, i.e.,
\bege
H|\vn{k}n\rangle= \mathcal{E}_{\vn{k}n} |\vn{k}n\rangle.
\ee
This simple form of $G^{\rm R}_{\vn{k}}(\mathcal{E})$ allows us to perform the
energy integrations in Eq.~\eqref{eq_chi_noabrev} analytically.
The resulting expressions are given in 
Appendix~\ref{sec_implementation}
for the case of zero temperature.

It is convenient to discuss the laser-induced torque $\vn{T}$ in terms
of the equivalent effective magnetic field $\vn{B}^{\rm eff}$ that one
needs to apply in order to produce the same torque on the
magnetization. It is given by
\bege\label{eq_def_eff_field}
\vn{B}^{\rm eff}=\frac{\vn{T}\times \hat{\vn{M}}}{\mu},
\ee
where $\mu$ is the magnetic moment in the simulation volume.

The expressions that we use to evaluate 
the non-equilibrium spin
density $\delta\vn{S}$, Eq.~\eqref{eq_define_delta_s}, 
are similar to Eq.~\eqref{eq_define_torque_chi} and 
Eq.~\eqref{eq_chi_noabrev}:
\bege\label{eq_define_spinpol_chi}
\delta S_{i}=-\frac{\hbar}{2}\frac{a_{0}^3 I}{c}
\frac{\mathcal{E}_{\rm H}}{
\left(
\hbar\omega
\right)^2
}
{\rm Im}
\sum_{jk}
\epsilon_j
\epsilon_k^*
\bar{\chi}_{ijk},
\ee
where
\bege\label{eq_chibar_noabrev}
\begin{aligned}
\bar{\chi}_{ijk}=&\frac{2}{\mathcal{N}\hbar a_0^2}\sum_{\vn{k}}
\int \rmd \mathcal{E}
{\rm Tr}\Big[\\
&f(\mathcal{E})
\sigma_{i}
G^{\rm R}_{\vn{k}}(\mathcal{E})
v_{j}
G^{\rm R}_{\vn{k}}(\mathcal{E}-\hbar\omega)
v_{k}
G^{\rm R}_{\vn{k}}(\mathcal{E})
\\
-&
f(\mathcal{E})
\sigma_{i}
G^R_{\vn{k}}(\mathcal{E})
v_{j}
G^R_{\vn{k}}(\mathcal{E}-\hbar\omega)
v_{k}
G^A_{\vn{k}}(\mathcal{E})\\
+&f(\mathcal{E})
\sigma_{i}
G^{\rm R}_{\vn{k}}(\mathcal{E})
v_{k}
G^{\rm R}_{\vn{k}}(\mathcal{E}+\hbar\omega)
v_{j}
G^{\rm R}_{\vn{k}}(\mathcal{E})
\\
-&f(\mathcal{E})
\sigma_{i}
G^{\rm R}_{\vn{k}}(\mathcal{E})
v_{k}
G^{\rm R}_{\vn{k}}(\mathcal{E}+\hbar\omega)
v_{j}
G^{\rm A}_{\vn{k}}(\mathcal{E})
\\
+&
f(\mathcal{E}-\hbar\omega)
\sigma_{i}
G^R_{\vn{k}}(\mathcal{E})
v_{j}
G^R_{\vn{k}}(\mathcal{E}-\hbar\omega)
v_{k}
G^A_{\vn{k}}(\mathcal{E})\\
+&
f(\mathcal{E}+\hbar\omega)
\sigma_{i}
G^{\rm R}_{\vn{k}}(\mathcal{E})
v_{k}
G^{\rm R}_{\vn{k}}(\mathcal{E}+\hbar\omega)
v_{j}
G^{\rm A}_{\vn{k}}(\mathcal{E})
\Big].\\
\end{aligned}
\ee

\section{Results}
\label{sec_results}
\subsection{Computational details}
\label{sec_computational_parameters}
We employ the full-potential linearized augmented-plane-wave (FLAPW)
program {\tt FLEUR}~\cite{fleurcode} in order to determine the
electronic structure of bcc Fe, $L1_0$ FePt and hcp Co selfconsistently
within the generalized-gradient 
approximation~\cite{PerdewBurkeEnzerhof} to
density-functional theory.
The experimental lattice constants are used.
In the case of Fe and FePt the crystallographic $c$ and $a$ axes are
aligned with the 
$z$ and $y$ directions, respectively (Figure 1 illustrates the coordinate frame).
In the case of Co we performed two calculations in order to assess the anisotropy
of the laser-induced torques: One calculation where 
the $c$ axis is aligned with the $z$ direction, and one where 
the $c$ axis is aligned with the $x$ direction (in both calculations the $a$ axis is
in $y$ direction). 

In order to perform the Brillouin zone integrations
in Eq.~\eqref{eq_chi_noabrev} and 
in Eq.~\eqref{eq_chibar_noabrev} 
computationally efficiently
based on the Wannier interpolation technique~\cite{rmp_wannier90},
we constructed 18 maximally localized Wannier functions (MLWFs) per
transition metal
atom from an $8\times 8\times 8$ $\vn{k}$~mesh~\cite{wannier90,WannierPaper}.
In order to describe room temperature experiments in Fe, FePt and Co,
it is a very good approximation to set the temperature in the Fermi
distribution function $f(\mathcal{E})$ in Eq.~\eqref{eq_chi_noabrev} and 
in Eq.~\eqref{eq_chibar_noabrev} to zero. Effects of room-temperature
phonon-scattering can be modelled by the phenomenological broadening
parameter $\Gamma$ in Eq.~\eqref{eq_define_green_analy}.
The energy
integrations in Eq.~\eqref{eq_chi_noabrev} and 
in Eq.~\eqref{eq_chibar_noabrev} 
are performed analytically, as described in 
Appendix~\ref{sec_implementation}. 
We vary $\Gamma$ in the range from 5~meV to 0.4~eV. For this range of
broadening $\Gamma$ we find that not more than
$256\times 256\times 256$ $\vn{k}$~points are needed in order
to converge the Brillouin zone sampling in
Eq.~\eqref{eq_chi_noabrev} and 
Eq.~\eqref{eq_chibar_noabrev}.

In section~\ref{sec_laser_induced_torques} and \ref{sec_laser_induced_spin} 
we discuss laser induced torques and
spin polarization for the laser intensity $I=10$~GW/cm$^{2}$.
The photon energy is set to 1.55~eV.
The light is propagating into the $x$ direction (as illustrated in Figure 1)
and the polarization vector is $\polarivec_{\lambda}=(0,1,i\lambda)/\sqrt{2}$, where
$\lambda=+1$ and $\lambda=-1$ describe left and right circularly polarized light, respectively.
The magnetization is set along the $z$ direction. 

\subsection{Laser-induced torques}
\label{sec_laser_induced_torques}
\begin{figure}
\includegraphics[width=\linewidth,trim=0.8cm 1cm 9cm 3cm,clip]{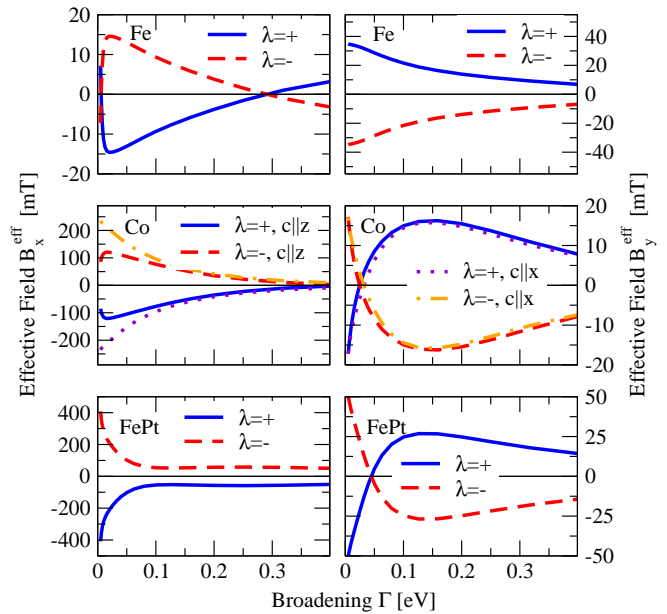}
\caption{\label{fig_xheli_torques_zmag}
Laser-induced effective magnetic
fields $B^{\rm eff}_{x}$ (left column) and $B^{\rm eff}_{y}$ (right column)
in Fe, Co and FePt
as a function of broadening $\Gamma$
at $I=10$~GW/cm$^{2}$. $\hat{\vn{M}}$
is in $z$ direction. $\lambda=+$ and $\lambda=-$ denote left
and right circularly polarized light, respectively. In the case of Co,
results are shown for $c$ axis in $z$ direction ($c\Vert z$) and
$c$ axis in $x$ direction ($c\Vert x$).
}
\end{figure}

We discuss the laser-induced torques in terms of the equivalent effective magnetic fields
defined in Eq.~\eqref{eq_def_eff_field}.
Figure~\ref{fig_xheli_torques_zmag} shows these laser-induced effective magnetic fields
in Fe, Co and FePt. 
In the case of Co we show results of two different calculations: One where the crystallographic $c$ axis 
is in $z$ direction ($c\Vert z$) and one where it is in $x$ direction ($c\Vert x$).
The effective field in $x$ direction, $B^{\rm eff}_{x}$, arises due to the
IFE in this case. The effective field in $y$ direction arises due to the OSTT.
Both $B^{\rm eff}_{x}$ and $B^{\rm eff}_{y}$ are odd in the helicity $\lambda$.
In the geometry of Figure 1 $B^{\rm eff}_{x}$ leads to an in-plane torque and thus an
initial in-plane tilting of the magnetization, while
$B^{\rm eff}_{y}$ leads to an out-of-plane torque and thus an initial out-of-plane tilting.
The effective fields depend strongly on the broadening $\Gamma$, which varies between
5~meV and 0.4~eV in the figure. In Fe $B^{\rm eff}_{y}$ is always 
larger than $B^{\rm eff}_{x}$ in the considered $\Gamma$-range, while in
FePt $B^{\rm eff}_{x}$ is always larger than $B^{\rm eff}_{y}$. In Co
$B^{\rm eff}_{x}$ dominates over $B^{\rm eff}_{y}$ for small and medium $\Gamma$, while
for very large broadening $B^{\rm eff}_{y}$ becomes larger than $B^{\rm eff}_{x}$.
In Co the component $B^{\rm eff}_{x}$ exhibits a strong anisotropy at small $\Gamma$.

In previous works we used $\Gamma$~=~25~meV to model room-temperature 
experiments on Co/Pt bilayers~\cite{ibcsoit}.
At $\Gamma$~=~25~meV we find $B^{\rm eff}_{x}=118$~mT
and $B^{\rm eff}_{y}=0.23$~mT
in Co for the $c\Vert z$ case.
For $c\Vert x$ we find $B^{\rm eff}_{x}=194$~mT
and $B^{\rm eff}_{y}=3.1$~mT in Co. Similarly large anisotropies have
been predicted for the anomalous Hall effect in Co~\cite{anisotropic_ahe}. 
At $\Gamma$~=~25~meV the component $B^{\rm eff}_{x}$ strongly dominates over
$B^{\rm eff}_{y}$, leading to an initial in-plane tilt of the magnetization in the geometry
of figure 1, consistent with the experimental 
interpretation~\cite{femtosecond_control_electric_currents_Huisman}.
For a 50~fs laser-pulse with a fluence 
of 1~mJ/cm$^2$~\cite{femtosecond_control_electric_currents_Huisman},
which corresponds to an intensity of the order 
of $I\approx1~$mJcm$^{-2}/(50$~fs)=~20~GWcm$^{-2}$,
an effective field of 200~mT in Co was estimated from experiments~\cite{femtosecond_control_electric_currents_Huisman}, corresponding to
roughly $100$~mT at $I$=~10~GWcm$^{-2}$. The experimental geometry corresponds 
to the $c\Vert x$ case in our simulation. Our theoretical result 
of $B^{\rm eff}_{x}=194$~mT is thus larger than
the experimental estimate by roughly a factor of 2. One potential reason for the
discrepancy is that laser-pulses were used in the experiment, while our simulation assumes
a continuous laser beam. Additionally, the effective magnetic field is strongly $\Gamma$ dependent
according to our calculation and any disorder present in the 10~nm Co film used in the experiment
might correspond to a value of $\Gamma$ larger than 25~meV, which we assumed in this comparison.

At $\Gamma$~=~25~meV $B^{\rm eff}_{x}$ strongly dominates over $B^{\rm eff}_{y}$ in Co and FePt.
On the other hand, the case of Fe shows that generally 
$B^{\rm eff}_{x}$ and $B^{\rm eff}_{y}$ can be of similar magnitude in
transition metal ferromagnets. If an Fe layer is used instead of the Co layer in Figure 1, 
the initial magnetization tilt will be a mixture of in-plane and out-of-plane according to
our calculations. While the 
helicity-dependent component of the photocurrent
in Co/Pt bilayers arises from
an initial in-plane tilting~\cite{femtosecond_control_electric_currents_Huisman}
combined with the odd component of the ISOT, also out-of-plane
tilting gives rise to photocurrents via the even ISOT component~\cite{invsot}. 
The photocurrent density $\vn{J}$ induced by the initial magnetization tilt 
in the bilayer geometry of figure~1 can be written 
as~\cite{femtosecond_control_electric_currents_Huisman}
\bege
\begin{aligned}
\vn{J}=&-\frac{\gamma t^{\rm odd}}{V}\hat{\vn{e}}_{x}
\times
\left[
\hat{\vn{M}}\times
\vn{B}^{\rm eff}
\right]\\
&-\frac{\gamma t^{\rm even}}{V}\hat{\vn{e}}_{x}
\times
\left[
\hat{\vn{M}}\times
\left(
\hat{\vn{M}}\times
\vn{B}^{\rm eff}
\right)
\right],
\end{aligned}
\ee
where $\gamma$ is the electron gyromagnetic factor, $V$ is the 
volume, $\hat{\vn{e}}_x$ is a unit
vector along the $x$ axis and the coefficients
$t^{\rm odd}$ and $t^{\rm even}$ characterize the odd and even component of the SOT, respectively.
When $\vn{B}^{\rm eff}$ points in $x$ direction, the photocurrent is proportional to $t^{\rm odd}$
and when $\vn{B}^{\rm eff}$ points in $y$ direction, the photocurrent is proportional to $t^{\rm even}$.
In both cases the photocurrent is flowing along the magnetization direction.
Therefore, we expect that the helicity-dependent component of
the photocurrent in experiments analogous to the ones in 
Ref.~\cite{femtosecond_control_electric_currents_Huisman} but based on
Fe/Pt bilayers contains contributions from both the even and odd ISOT. 
The differences in the effective fields $\vn{B}^{\rm eff}$ between
Fe, Co and FePt suggest that ferromagnetic materials can be designed such that the IFE is zero and
the OSTT is nonzero. Using such materials in experiments analogous to the ones in 
Ref.~\cite{femtosecond_control_electric_currents_Huisman} would allow 
the contactless measurement of the even ISOT,
which contains information about the spin Hall effect, 
from the helicity-odd component of the photocurrent. In fact, the
helicity-even
component of the photocurrent is already used for contactless
measurement of the spin Hall 
effect~\cite{thz_emitter_Seifert}.  

\begin{figure}
\includegraphics[width=\linewidth,trim=0.8cm 12cm 9cm 3cm,clip]{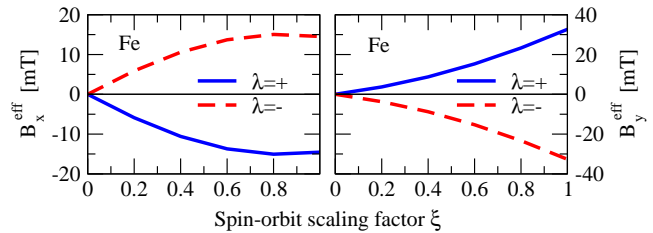}
\caption{\label{fig_torque_Fe_socscale}
Laser-induced effective magnetic 
fields $B^{\rm eff}_{x}$ (left) and $B^{\rm eff}_{y}$ (right) 
in Fe 
as a function of SOI scaling factor $\xi$
at $I=10$~GW/cm$^{2}$. $\hat{\vn{M}}$
is in $z$ direction. $\lambda=+$ and $\lambda=-$ denote left
and right circularly polarized light, respectively.
}
\end{figure}

In order to investigate the dependence of $B^{\rm eff}_{x}$ and $B^{\rm eff}_{y}$
on SOI, we linearly scale the spin-orbit interaction in the Hamiltonian with a factor $\xi$
such that SOI is switched off for $\xi$=0 and that the full SOI is active for $\xi$=1. 
Figure~\ref{fig_torque_Fe_socscale} shows the laser-induced effective magnetic fields
in Fe as a function of $\xi$. When SOI is switched off $B^{\rm eff}_{x}$ and $B^{\rm eff}_{y}$
vanish, which proves that SOI is the origin of these laser-induced effective magnetic fields.

\subsection{Laser-induced spin polarization}
\label{sec_laser_induced_spin}
\begin{figure}
\includegraphics[width=\linewidth,trim=0.8cm 1cm 9cm 3cm,clip]{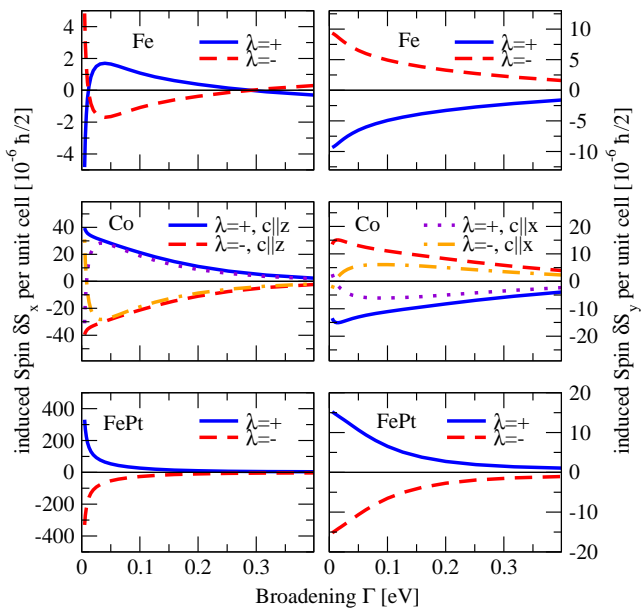}
\caption{\label{fig_xheli_spinpol}
Laser-induced spin 
polarization $\delta S_{x}$ (left column) and $\delta S_{y}$ (right column) 
in Fe, Co and FePt 
as a function of broadening $\Gamma$
at $I=10$~GW/cm$^{2}$. $\hat{\vn{M}}$
is in $z$ direction. $\lambda=+$ and $\lambda=-$ denote left
and right circularly polarized light, respectively. In the case of Co,
results are shown for $c$ axis in $z$ direction ($c\Vert z$) and
$c$ axis in $x$ direction ($c\Vert x$).
}
\end{figure}

We first discuss the two components of the laser-induced spin polarization $\delta \vn{S}$
that are perpendicular to the magnetization, which points in 
$z$ direction. These perpendicular components are 
expected to be related to $B^{\rm eff}_{x}$ and $B^{\rm eff}_{y}$ discussed in the
previous section. 
Figure~\ref{fig_xheli_spinpol} shows that both $\delta S_{x}$ and $\delta S_{y}$ 
are odd in the helicity $\lambda$.
Due to Eq.~\eqref{eq_torque_from_spin_density} we expect similarities between
$B^{\rm eff}_{x}$ (Figure~\ref{fig_xheli_torques_zmag})
and $\delta S_{x}$
and between $B^{\rm eff}_{y}$ and $\delta S_{y}$.
Indeed, in Fe $\delta S_{j}$  exhibits the same
qualitative dependence on $\Gamma$ as its counterpart $B^{\rm eff}_{j}$ ($j=x,y$). 
Since the electron spin magnetic moment 
is antiparallel to the electron 
spin, $\delta S_{j}$ and $B^{\rm eff}_{j}$ are opposite in sign for a given helicity $\lambda$.
In FePt only $B^{\rm eff}_{x}$ and $\delta S_{x}$ behave similarly as a function of $\Gamma$,
while $B^{\rm eff}_{y}$ and $\delta S_{y}$ exhibit different trends, notably a sign change in
$B^{\rm eff}_{y}$ that is absent in $\delta S_{y}$. In Co both $\delta S_{x}$ and $\delta S_{y}$
are strongly anisotropic for small $\Gamma$, while only $B^{\rm eff}_{x}$ displays strong anisotropy.
These qualitative differences between $\delta S_{j}$ and $B^{\rm eff}_{j}$ illustrate the importance
of calculating the torques and effective magnetic fields from Eq.~\eqref{eq_torque_from_spin_density}, which
takes into account that the exchange field varies strongly on the atomic scale.
On the other hand, in Fe, where $\delta S_{j}$ and $B^{\rm eff}_{j}$ behave very similarly, it is tempting to
define an effective exchange field $\Bxc^{\rm xc}_{\rm eff}$
by the equation
\bege\label{eq_estimate_torque}
\vn{T}=\frac{2\mu_{\rm B}}{\hbar}\Bxc^{\rm xc}_{\rm eff}\delta\vn{S}\times\hat{\vn{M}}.
\ee
The corresponding exchange splitting is 
\bege
\Delta V_{\rm eff}=2\mu_{\rm B}\Bxc^{\rm xc}_{\rm eff}=-\frac{\hbar\mu B^{\rm eff}_{j}}{\delta S_{j}},
\ee
where $\mu$ is the magnetic moment per unit cell. 
From our results of $B^{\rm eff}_{j}$ and $\delta S_{j}$ in Fe at $\Gamma$~=~25~meV we obtain
$\Delta V_{\rm eff}$~=~2.6~eV for $j=1$ and $\Delta V_{\rm eff}$~=~1.1~eV for $j=2$. The finding
that we obtain different values for $j=1$ and $j=2$ shows that Eq.~\eqref{eq_estimate_torque}
can not be used for precise calculations in Fe.
However, since $\Delta V_{\rm eff}$ has the expected order of magnitude 
of the exchange splitting in Fe, one can indeed
use Eq.~\eqref{eq_estimate_torque} for rough estimates of the torque $\vn{T}$ from the induced
spin polarization $\delta \vn{S}$ in certain cases. 

\begin{figure}
\includegraphics[width=\linewidth,trim=0.0cm 7cm 10cm 3cm,clip]{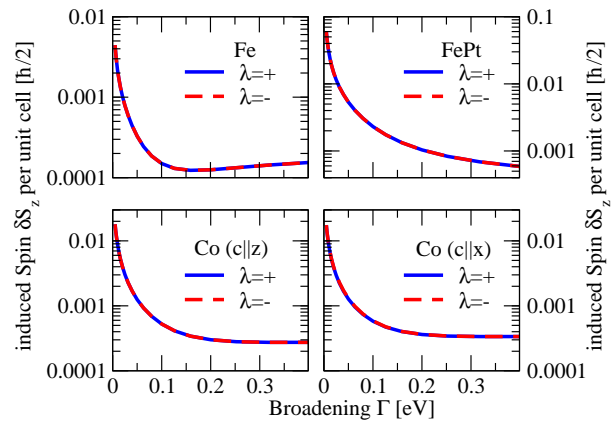}
\caption{\label{fig_xheli_spinpolparallel}
Laser-induced spin 
polarization $\delta S_{z}$ 
in Fe, Co and FePt 
as a function of broadening $\Gamma$
at $I=10$~GW/cm$^{2}$. $\hat{\vn{M}}$
is in $z$ direction. $\lambda=+$ and $\lambda=-$ denote left
and right circularly polarized light, respectively. In the case of Co,
results are shown for $c$ axis in $z$ direction ($c\Vert z$) and
$c$ axis in $x$ direction ($c\Vert x$).
}
\end{figure}

Next, we discuss the laser-induced spin polarization $\delta S_{z}$
along the magnetization direction, which is shown 
in Figure~\ref{fig_xheli_spinpolparallel}. We find $\delta S_{z}$ to be
helicity-independent, in agreement with recent calculations based on the
Liouville von-Neumann equation~\cite{ife_berritta}. 
Interestingly, $\delta S_{z}$ reaches much larger values than 
the two perpendicular components $\delta S_{x}$
and $\delta S_{y}$. For example in 
FePt 
at $\Gamma$~=~25~meV we find
$\delta S_{z}$~=~1.2$\cdot 10^{-2}\hbar/2$
compared to only
$\delta S_{x}$~=~9.2$\cdot 10^{-5}\hbar/2$
and
$\delta S_{y}$~=~1.3$\cdot 10^{-5}\hbar/2$.
In the case of Co $\delta S_{z}$ depends on whether the $c$ axis is in $x$ or $z$ direction,
but this anisotropy is less striking than for $\delta S_{x}$ and $\delta S_{y}$ at small $\Gamma$.

\section{Summary}
\label{sec_summary}
We combine \textit{ab-initio} electronic structure calculations
with the Keldysh nonequilibrium formalism
in order to study laser-induced
torques and nonequilibrium spin polarization in bcc Fe, hcp Co 
and $L1_0$
FePt. 
Our calculations show that both IFE and OSTT are nonzero in these
metallic ferromagnets.
In the case of Fe the torques due to the OSTT are larger
than those due to the IFE, in FePt the IFE dominates over the OSTT and in Co
the IFE is dominant only for small and medium quasiparticle broadenings.
In view of this strong dependence of the IFE/OSTT ratio on the
ferromagnetic material and the quasiparticle broadening (and hence the disorder in
the system) it should be possible to design materials such that they
display either IFE-torques or OSTT, but not both at the same
time. This allows the contactless measurement of various spintronics
effect in optical experiments. 
We find the torques and the perpendicular component of the 
nonequilibrium spin polarization to be odd in the helicity of the
laser light, while the spin polarization that is induced parallel to
the
magnetization is helicity-independent. This parallel component of the
nonequilibrium spin polarization can be orders of magnitude larger than the
perpendicular component.
The comparison between laser-induced torques and laser-induced
nonequilibrium
spin density shows the importance of using the torque operator 
for calculations of laser-induced torques in realistic materials in
order to capture the variation of the exchange field on the atomic
scale.
We find that both the laser-induced torques and the laser-induced
nonequilibrium spin polarization are anisotropic in Co. 
In the case of hcp Co we find
good agreement between the calculated laser-induced torque and a
recent experiment.

\appendix
\section{Formalism}
\label{sec_formalism}
The Green function $G$ in the presence of the perturbing laser field
is obtained from the unperturbed Green 
function $G_{\rm eq}$ via  
the Dyson equation on the Keldysh contour~\cite{rammer_smith}
\bege\label{eq_dyson_keldysh}
G(1,1')
\!=\!
G_{\rm eq}(1,1')
\!+\!
\!
\int\!\! \rmd\,2\, G_{\rm eq}(1,2) 
\frac{\delta H(2)}{\hbar} 
G(2,1'),
\ee
where $\delta H$ is the 
perturbation Eq.~\eqref{eq_laser_perturbation} 
due to the electric field of the laser.
We iterate Eq.~\eqref{eq_dyson_keldysh} to obtain a power series in $\delta H$
and identify the term quadratic in $\delta H$. Applying the
Langreth theorem
\bege
(GGG)^{<}=G^{\rm R}G^{\rm R}G^{<}
+G^{\rm R}G^{<}G^{\rm A}+
G^{<}G^{\rm A}G^{\rm A}
\ee
to the term quadratic in $\delta H$ we obtain:
\bege\label{eq_dy_secord}
\begin{aligned}
&G^{<}_{2}(t,t')=\\
&\int\!\!\rmd t_1
\!\!\int\!\!\rmd t_2\, 
G^{\rm R}_{\rm eq}(t,t_1)
\frac{\delta H(t_1)}{\hbar}
G^{\rm R}_{\rm eq}(t_1,t_2)
\frac{\delta H(t_2)}{\hbar}
G^{<}_{\rm eq}(t_2,t')+\\
&
\int\!\!\rmd t_1
\!\!\int\!\!\rmd t_2\, 
G^{\rm R}_{\rm eq}(t,t_1)
\frac{\delta H(t_1)}{\hbar}
G^{<}_{\rm eq}(t_1,t_2)
\frac{\delta H(t_2)}{\hbar}
G_{\rm eq}^{\rm A}(t_2,t')+\\
&
\int\!\!\rmd t_1
\!\!\int\!\!\rmd t_2\,
G_{\rm eq}^{<}(t,t_1)
\frac{\delta H(t_1)}{\hbar}
G_{\rm eq}^{\rm A}(t_1,t_2)
\frac{\delta H(t_2)}{\hbar}
G_{\rm eq}^{\rm A}(t_2,t').
\end{aligned}
\ee
Using
\bege
G^{\rm R}_{\rm eq}(t,t')=\frac{1}{2\pi\hbar}
\int_{-\infty}^{\infty}
\!\!\rmd \mathcal{E}\, \rme^{-i\mathcal{E}(t-t')/\hbar}G_{\rm eq}^{\rm R}(\mathcal{E})
\ee
and
\bege
\begin{aligned}
&\int_{-\infty}^{\infty}
\!\!\rmd t_1
\int_{-\infty}^{\infty}
\!\!\rmd t_2\,
\rme^{-\frac{i}{\hbar}\mathcal{E}(t-t_1)}
e^{-i\omega_1 t_1}
\rme^{-\frac{i}{\hbar}\mathcal{E}'(t_1-t_2)}\times\\
&\quad\quad\quad\quad\quad\quad\quad\quad\quad\quad\times e^{-i\omega_2 t_2}
\rme^{-\frac{i}{\hbar}\mathcal{E}''(t_2-t')}
=\\
&=h^2
\delta(\mathcal{E}\!-\!\hbar\omega_1\!-\!\mathcal{E}'  )
\delta(\mathcal{E}'\!-\!\hbar\omega_2\!-\!\mathcal{E}''  )
\rme^{-\frac{i}{\hbar}\mathcal{E}t}
\rme^{\frac{i}{\hbar}\mathcal{E}''t'}
\end{aligned}
\ee
the time-integration of the product of three Green functions can be performed easily:
\bege
\begin{aligned}
&\int\!\!\!
 \rmd t_1\!\!\!
\int\!\!\!
\rmd t_2
G_{\rm eq}^{\alpha}(t,t_1)
e^{-i\omega_1 t_1}
G_{\rm eq}^{\alpha'}(t_1,t_2)
e^{-i\omega_2 t_2}
G_{\rm eq}^{\alpha''}(t_2,t)
=\\
&=
\frac{\rme^{-i[\omega_1+\omega_2]t}}{h}
\!\int\!\!
\rmd \mathcal{E}
G_{\rm eq}^{\alpha}(\mathcal{E}\!+\!\hbar\omega_1)
G_{\rm eq}^{\alpha'}(\mathcal{E})
G_{\rm eq}^{\alpha''}\!(\mathcal{E}\!-\!\hbar\omega_2),\\
\end{aligned}
\ee
where $\alpha,\alpha',\alpha''={\rm R,A,<}$ and $\omega_1,\omega_2=\pm\omega$.
As discussed in section~\ref{sec_computational_method}, we only need the
dc component of $G^{<}_2$, which arises from all terms with $\omega_1=-\omega_2=\pm\omega$.
It is given by
\bege\label{eq_lesser_dc_noabrev}
\begin{aligned}
G^{<}_{\rm dc}&=
\frac{e^2E^2_{0}}{8\pi\omega^2\hbar^3}\int \rmd \mathcal{E}\Big\{\\
&
G_{\rm eq}^{\rm R}(\mathcal{E}-\hbar\omega)
\vn{v}\cdot\polarivec^{*}
G_{\rm eq}^{\rm R}(\mathcal{E})
\vn{v}\cdot\polarivec
G_{\rm eq}^{<}(\mathcal{E}-\hbar\omega)+\\
+&G_{\rm eq}^{\rm R}(\mathcal{E}+\hbar\omega)
\vn{v}\cdot\polarivec
G_{\rm eq}^{\rm R}(\mathcal{E})
\vn{v}\cdot\polarivec^{*}
G_{\rm eq}^{<}(\mathcal{E}+\hbar\omega)+\\
+&
G_{\rm eq}^{\rm R}(\mathcal{E}-\hbar\omega)
\vn{v}\cdot\polarivec^{*}
G_{\rm eq}^{<}(\mathcal{E})
\vn{v}\cdot\polarivec
G_{\rm eq}^{\rm A}(\mathcal{E}-\hbar\omega)+\\
+&G_{\rm eq}^{\rm R}(\mathcal{E}+\hbar\omega)
\vn{v}\cdot\polarivec
G_{\rm eq}^{<}(\mathcal{E})
\vn{v}\cdot\polarivec^{*}
G_{\rm eq}^{\rm A}(\mathcal{E}+\hbar\omega)+\\
+&
G_{\rm eq}^{<}(\mathcal{E}-\hbar\omega)
\vn{v}\cdot\polarivec^{*}
G_{\rm eq}^{\rm A}(\mathcal{E})
\vn{v}\cdot\polarivec
G_{\rm eq}^{\rm A}(\mathcal{E}-\hbar\omega)+\\
+&G_{\rm eq}^{<}(\mathcal{E}+\hbar\omega)
\vn{v}\cdot\polarivec
G_{\rm eq}^{\rm A}(\mathcal{E})
\vn{v}\cdot\polarivec^{*}
G_{\rm eq}^{\rm A}(\mathcal{E}+\hbar\omega)\Big\}.
\end{aligned}
\ee
Substituting Eq.~\eqref{eq_lesser_dc_noabrev} into 
Eq.~\eqref{eq_torque_from_lesser} and using
\bege
G^{<}_{\rm eq}(\mathcal{E})=f(\mathcal{E})
\left[
G^{\rm A}_{\rm eq}(\mathcal{E})
-
G^{\rm R}_{\rm eq}(\mathcal{E})
\right]
\ee
yields
\bege\label{eq_ifator_dc2_shorthand}
\begin{aligned}
\vn{T}=&
\frac{ie^2E^2_{0}}{8\pi\omega^2\hbar^3}\int \rmd \mathcal{E}f(\mathcal{E})
{\rm Tr}\Big\{\\
&
\left[
\vn{\mathcal{T}}
R
\Lambda^{\dagger}
R^{+}
\Lambda
A
\right]-\left[\vn{\mathcal{T}}
R
{\Lambda}^{\dagger}
R^{+}
{\Lambda}
{R}
\right]
+\\
+&
\left[
\vn{\mathcal{T}}
{R}
{\Lambda}
{R}^{-}
{\Lambda}^{\dagger}
{A}
\right]-
\left[
\vn{\mathcal{T}}
{R}
{\Lambda}
{R}^{-}
{\Lambda}^{\dagger}
{R}
\right]+\\
+&
\left[
\vn{\mathcal{T}}
{R}^{-}
{\Lambda}^{\dagger}
{A}
{\Lambda}
{A}^{-}
\right]-
\left[
\vn{\mathcal{T}}
{R}^{-}
{\Lambda}^{\dagger}
{R}
{\Lambda}
{A}^{-}
\right]+\\
+&\left[
\vn{\mathcal{T}}
{R}^+
{\Lambda}
{A}
{\Lambda}^{\dagger}
{A}^+
\right]-
\left[
\vn{\mathcal{T}}
{R}^+
{\Lambda}
{R}
{\Lambda}^{\dagger}
{A}^+
\right]+\\
+&\left[
\vn{\mathcal{T}}
{A}
{\Lambda}^{\dagger}
{A}^+
{\Lambda}
{A}
\right]-\left[
\vn{\mathcal{T}}
{R}
{\Lambda}^{\dagger}
{A}^+
{\Lambda}
{A}
\right]+\\
+&
\left[
\vn{\mathcal{T}}
{A}
{\Lambda}
{A}^{-}
{\Lambda}^{\dagger}
{A}
\right]-
\left[
\vn{\mathcal{T}}
{R}
{\Lambda}
{A}^-
{\Lambda}^{\dagger}
{A}
\right]
\Big\},
\end{aligned}
\ee
where we introduced the abbreviations
$
\Lambda=\vn{v}\cdot\polarivec
$,
$
\Lambda^{\dagger}=
\left[
\vn{v}\cdot\polarivec
\right]^{\dagger}=\vn{v}\cdot\polarivec^*
$,
$
R=G^{\rm R}_{\rm eq}(\mathcal{E})
$,
$
A=G^{\rm A}_{\rm eq}(\mathcal{E})
$,
$
R^+=G^{\rm R}_{\rm eq}(\mathcal{E}+\hbar\omega)
$,
$
A^+=G^{\rm A}_{\rm eq}(\mathcal{E}+\hbar\omega)
$,
$
R^-=G^{\rm R}_{\rm eq}(\mathcal{E}-\hbar\omega)
$
and
$
A^-=G^{\rm A}_{\rm eq}(\mathcal{E}-\hbar\omega)
$.
Terms that contain more than one $A$ can be rewritten
as complex conjugates of terms with more than one $R$:
\bege\label{eq_ifator_dc2_cc}
\begin{aligned}
&\vn{T}=
\frac{ie^2E^2_{0}}{8\pi\omega^2 \hbar^3}\int \rmd \mathcal{E}f(\mathcal{E})
{\rm Tr}\Big\{\\
&
\left[
\vn{\mathcal{T}}
R
\Lambda^{\dagger}
R^{+}
\Lambda
A
\right]-\left[\vn{\mathcal{T}}
R
{\Lambda}^{\dagger}
R^{+}
{\Lambda}
{R}
\right]
+\\
+&
\left[
\vn{\mathcal{T}}
{R}
{\Lambda}
{R}^{-}
{\Lambda}^{\dagger}
{A}
\right]-
\left[
\vn{\mathcal{T}}
{R}
{\Lambda}
{R}^{-}
{\Lambda}^{\dagger}
{R}
\right]+\\
+&
\left[
\vn{\mathcal{T}}
{R}^{-}
{\Lambda}^{\dagger}
{R}
{\Lambda}
{A}^{-}
\right]^{*}-
\left[
\vn{\mathcal{T}}
{R}^{-}
{\Lambda}^{\dagger}
{R}
{\Lambda}
{A}^{-}
\right]+\\
+&\left[
\vn{\mathcal{T}}
{R}^{+}
{\Lambda}
{R}
{\Lambda}^{\dagger}
{A}^{+}
\right]^{*}-
\left[
\vn{\mathcal{T}}
{R}^{+}
{\Lambda}
{R}
{\Lambda}^{\dagger}
{A}^{+}
\right]
+\\
+&\left[
\vn{\mathcal{T}}
{R}
{\Lambda}^{\dagger}
{R}^+
{\Lambda}
{R}
\right]^{*}
-\left[
\vn{\mathcal{T}}
{R}
{\Lambda}^{\dagger}
{R}^+
{\Lambda}
{A}
\right]^{*}+\\
+&
\left[
\vn{\mathcal{T}}
{R}
{\Lambda}
{R}^{-}
{\Lambda}^{\dagger}
{R}
\right]^{*}-
\left[
\vn{\mathcal{T}}
{R}
{\Lambda}
{R}^-
{\Lambda}^{\dagger}
{A}
\right]^{*}
\Big\}.
\end{aligned}
\ee
Using the imaginary part to simplify the expression 
and introducing a Brillouin zone average over $\mathcal{N}$
$\vn{k}$ points we finally obtain
\bege
\begin{aligned}
\vn{T}=&\frac{|e|^2E^2_{0}}{4\pi\omega^2\hbar^3\mathcal{N}}\sum_{\vn{k}}\int \rmd \mathcal{E}
{\rm Im}{\rm Tr}\Big\{\\
&f(\mathcal{E})
\left[
\vn{\mathcal{T}}
R_{\vn{k}}
\Lambda
R^{-}_{\vn{k}}
\Lambda^{\dagger}
R_{\vn{k}}
+
\vn{\mathcal{T}}
R_{\vn{k}}
\Lambda^{\dagger}
R^{+}_{\vn{k}}
\Lambda
R_{\vn{k}}
\right]\\
+&
\left[
f(\mathcal{E}-\hbar\omega)
-
f(\mathcal{E})
\right]
\left[
\vn{\mathcal{T}}
R_{\vn{k}}
{\Lambda}
R^{-}_{\vn{k}}
{\Lambda}^{\dagger}
{A}_{\vn{k}}
\right]
+\\
+&
\left[
f(\mathcal{E}+\hbar\omega)
-
f(\mathcal{E})
\right]
\left[
\vn{\mathcal{T}}
{R}_{\vn{k}}
{\Lambda}^{\dagger}
{R}^{+}_{\vn{k}}
{\Lambda}
{A}_{\vn{k}}
\right]
\Big\}\\
=&\frac{a_{0}^3 I}{c}
\left(
\frac{\mathcal{E}_{\rm H}}{\hbar\omega}
\right)^2
{\rm Im}
\sum_{ijk}\hat{\vn{e}}_{i}
(\hat{\vn{e}}_{j}\cdot\polarivec)
(\hat{\vn{e}}_{k}\cdot\polarivec^*)
\chi_{ijk},
\end{aligned}
\ee
where 
$\hat{\vn{e}}_{1}$, $\hat{\vn{e}}_{2}$ and $\hat{\vn{e}}_{3}$ are
unit vectors along the $x$, $y$ and $z$ axes, respectively.
The coefficient $\chi_{ijk}=\chi^{(1)}_{ijk}+\chi^{(2)}_{ijk}$ is given by
\bege\label{eq_chi1}
\begin{aligned}
\chi^{(1)}_{ijk}=&\frac{2}{\mathcal{N}\hbar a_0^2 \mathcal{E}_{\rm H}}\sum_{\vn{k}}
\int \rmd \mathcal{E}
{\rm Tr}\Big\{\\
&
\left[
f(\mathcal{E}-\hbar\omega)
-
f(\mathcal{E})
\right]
\left[
\mathcal{T}_{i}
R_{\vn{k}}
v_{j}
R^{-}_{\vn{k}}
v_{k}
A_{\vn{k}}
\right]
+\\
+&
\left[
f(\mathcal{E}+\hbar\omega)
-
f(\mathcal{E})
\right]
\left[
\mathcal{T}_{i}
{R}_{\vn{k}}
v_{k}
{R}^{+}_{\vn{k}}
v_{j}
{A}_{\vn{k}}
\right]
\Big\}\\
\end{aligned}
\ee
and
\bege\label{eq_chi2}
\begin{aligned}
\chi^{(2)}_{ijk}=&\frac{2}{\mathcal{N}\hbar a_0^2 \mathcal{E}_{\rm H}}\sum_{\vn{k}}
\int \rmd \mathcal{E}
f(\mathcal{E})\\
&{\rm Tr}
\left[
\mathcal{T}_{i}
R_{\vn{k}}
v_{j}
R^{-}_{\vn{k}}
v_{k}
R_{\vn{k}}
+
\mathcal{T}_{i}
R_{\vn{k}}
v_{k}
R^{+}_{\vn{k}}
v_{j}
R_{\vn{k}}
\right].
\end{aligned}
\ee

\section{Expressions at $T=0$K}
\label{sec_implementation}
In the present paper we use 
the constant broadening $\Gamma$
in order to simulate disorder and finite lifetimes of the electronic states.
Therefore, the energy dependence of the Green function
is known analytically:
\bege
R_{\vn{k}}=G^{\rm R}_{\vn{k}}(\mathcal{E})=
\hbar
\sum_{n}\frac{|\vn{k}n\rangle\langle\vn{k}n|}{\mathcal{E}-\mathcal{E}_{\vn{k}n}+i\Gamma}.
\ee
This simple form of $G^{\rm R}_{\vn{k}}(\mathcal{E})$ allows us to perform the
energy integrations in Eq.~\eqref{eq_chi1} and Eq.~\eqref{eq_chi2} analytically.
We discuss only the zero-temperature limit and therefore replace the Fermi function by the Heaviside
step function as $f(\mathcal{E})=\theta(\mathcal{E}_{\rm F}-\mathcal{E})$, where
$\mathcal{E}_{\rm F}$ is the Fermi energy. 
Thus, we need the following two integrals for the 
evaluation of Eq.~\eqref{eq_chi1} and Eq.~\eqref{eq_chi2} in the zero-temperature limit:
\bege\label{eq_integ_i1}
\begin{aligned}
&I_{1}(\mathcal{E}_1,\mathcal{E}_2,\mathcal{E}_3,\mathcal{E}_{4})=\\
&=\int_{-\infty}^{\mathcal{E}_{4}}
\frac{\mathcal{E}^2_{\rm H}\quad\rmd\mathcal{E}}{
(\mathcal{E}-\mathcal{E}_1+i\Gamma)
(\mathcal{E}-\mathcal{E}_2+i\Gamma)
(\mathcal{E}-\mathcal{E}_3+i\Gamma)
}
\end{aligned}
\ee
and
\bege\label{eq_integ_i2}
\begin{aligned}
&I_{2}(\mathcal{E}_1,\mathcal{E}_2,\mathcal{E}_3,\mathcal{E}_{4})=\\
&=\int_{-\infty}^{\mathcal{E}_{4}}
\frac{ \mathcal{E}^2_{\rm H}\quad \rmd\mathcal{E}}{
(\mathcal{E}-\mathcal{E}_1+i\Gamma)
(\mathcal{E}-\mathcal{E}_2+i\Gamma)
(\mathcal{E}-\mathcal{E}_3-i\Gamma)
}.
\end{aligned}
\ee
In terms of $I_{1}(\mathcal{E}_1,\mathcal{E}_2,\mathcal{E}_3,\mathcal{E}_{4})$ 
and $I_{2}(\mathcal{E}_1,\mathcal{E}_2,\mathcal{E}_3,\mathcal{E}_{4})$
the coefficients $\chi^{(1)}_{ijk}$ and $\chi^{(2)}_{ijk}$ can be expressed as follows:
\bege\label{eq_chi1_analy}
\begin{aligned}
\chi^{(1)}_{ijk}=&\frac{2}{\mathcal{N}}\sum_{\vn{k}nmm'}
{\rm Im}\Bigl\{\\
\Bigl[
&I_{2}(\mathcal{E}_{\vn{k}m},\mathcal{E}_{\vn{k}m'}+\hbar\omega,\mathcal{E}_{\vn{k}n},\mathcal{E}_{\rm F}+\hbar\omega)
-\\
&I_{2}(\mathcal{E}_{\vn{k}m},\mathcal{E}_{\vn{k}m'}+\hbar\omega,\mathcal{E}_{\vn{k}n},\mathcal{E}_{\rm F})
\Bigl]
\mathcal{M}_{ijk}^{nmm'}+\\
\Bigl[
&I_{2}(\mathcal{E}_{\vn{k}m},\mathcal{E}_{\vn{k}m'}-\hbar\omega,\mathcal{E}_{\vn{k}n},\mathcal{E}_{\rm F}-\hbar\omega)
-\\
&I_{2}(\mathcal{E}_{\vn{k}m},\mathcal{E}_{\vn{k}m'}-\hbar\omega,\mathcal{E}_{\vn{k}n},\mathcal{E}_{\rm F})
\Bigl]
\mathcal{M}_{ikj}^{nmm'}
\Bigr\}
\end{aligned}
\ee
and
\bege\label{eq_chi2_analy}
\begin{aligned}
\chi^{(2)}_{ijk}=&\frac{2}{\mathcal{N}}\sum_{\vn{k}nmm'}
{\rm Im}\Bigl\{\\
&I_{1}(\mathcal{E}_{\vn{k}m},\mathcal{E}_{\vn{k}m'}+\hbar\omega,\mathcal{E}_{\vn{k}n},\mathcal{E}_{\rm F})
\mathcal{M}_{ijk}^{nmm'}+\\
&I_{1}(\mathcal{E}_{\vn{k}m},\mathcal{E}_{\vn{k}m'}-\hbar\omega,\mathcal{E}_{\vn{k}n},\mathcal{E}_{\rm F})
\mathcal{M}_{ikj}^{nmm'}
\Bigr\}
\end{aligned}
\ee
where
\bege
\mathcal{M}_{ijk}^{nmm'}=
\frac{
\langle
\vn{k}n
|
\mathcal{T}_{i}
|
\vn{k}m
\rangle
\langle
\vn{k}m
|
v_{j}
|
\vn{k}m'
\rangle
\langle
\vn{k}m'
|
v_{k}
|
\vn{k}n
\rangle}{
\mathcal{E}_{\rm H}
\left[
a_0 \frac{\mathcal{E}_{\rm H}}{\hbar}    
\right]^2
}.
\ee
The integrations in Eq.~\eqref{eq_integ_i1} and Eq.~\eqref{eq_integ_i2} can be performed analytically.
In the general case of $\mathcal{E}_1\ne\mathcal{E}_2\ne\mathcal{E}_3\ne\mathcal{E}_1$ we obtain
\bege\label{eq_rrr_alldiff}
\begin{aligned}
&I_{1}(\mathcal{E}_1,\mathcal{E}_2,\mathcal{E}_3,\mathcal{E}_{4})=\\
&
\frac{\mathcal{E}^2_{\rm H}}{2(\mathcal{E}_1-\mathcal{E}_2)(\mathcal{E}_1-\mathcal{E}_3)}
\log
\left[
1+\frac{(\mathcal{E}_1-\mathcal{E}_4)^2}{\Gamma^2}
\right]\\
+&
\frac{\mathcal{E}^2_{\rm H}}{2(\mathcal{E}_2-\mathcal{E}_3)(\mathcal{E}_2-\mathcal{E}_1)}
\log
\left[
1+\frac{(\mathcal{E}_2-\mathcal{E}_4)^2}{\Gamma^2}
\right]\\
+&
\frac{\mathcal{E}^2_{\rm H}}{2(\mathcal{E}_3-\mathcal{E}_1)(\mathcal{E}_3-\mathcal{E}_2)}
\log
\left[
1+\frac{(\mathcal{E}_3-\mathcal{E}_4)^2}{\Gamma^2}
\right]\\
+&
\frac{\mathcal{E}^2_{\rm H}}{i(\mathcal{E}_1-\mathcal{E}_2)(\mathcal{E}_1-\mathcal{E}_3)}
\arctan
\frac{\mathcal{E}_4-\mathcal{E}_1}{\Gamma}
\\
+&
\frac{\mathcal{E}^2_{\rm H}}{i(\mathcal{E}_2-\mathcal{E}_3)(\mathcal{E}_2-\mathcal{E}_1)}
\arctan
\frac{\mathcal{E}_4-\mathcal{E}_2}{\Gamma}
\\
+&
\frac{\mathcal{E}^2_{\rm H}}{i(\mathcal{E}_3-\mathcal{E}_1)(\mathcal{E}_3-\mathcal{E}_2)}
\arctan
\frac{\mathcal{E}_4-\mathcal{E}_3}{\Gamma}
\end{aligned}
\ee
and
\bege\label{eq_rra_alldiff}
\begin{aligned}
&I_{2}(\mathcal{E}_1,\mathcal{E}_2,\mathcal{E}_3,\mathcal{E}_{4})=\\
&
\frac{\mathcal{E}^2_{\rm H}}{2(\mathcal{E}_1-\mathcal{E}_2)(\mathcal{E}_1-\mathcal{E}_3-2i\Gamma)}
\log
\left[
1+\frac{(\mathcal{E}_1-\mathcal{E}_4)^2}{\Gamma^2}
\right]\\
+&
\frac{\mathcal{E}^2_{\rm H}}{2(\mathcal{E}_2-\mathcal{E}_3-2i\Gamma)(\mathcal{E}_2-\mathcal{E}_1)}
\log
\left[
1+\frac{(\mathcal{E}_2-\mathcal{E}_4)^2}{\Gamma^2}
\right]\\
+&
\frac{\mathcal{E}^2_{\rm H}}{2(\mathcal{E}_3-\mathcal{E}_1+2i\Gamma)(\mathcal{E}_3-\mathcal{E}_2+2i\Gamma)}
\log
\left[
1+\frac{(\mathcal{E}_3-\mathcal{E}_4)^2}{\Gamma^2}
\right]\\
+&
\frac{i\mathcal{E}^2_{\rm H}}{(\mathcal{E}_1-\mathcal{E}_2)(\mathcal{E}_3-\mathcal{E}_1+2i\Gamma)}
\left[
\frac{\pi}{2}+
\arctan
\frac{\mathcal{E}_4-\mathcal{E}_1}{\Gamma}
\right]\\
+&
\frac{i\mathcal{E}^2_{\rm H}}{(\mathcal{E}_3-\mathcal{E}_2+2i\Gamma)(\mathcal{E}_2-\mathcal{E}_1)}
\left[
\frac{\pi}{2}+
\arctan
\frac{\mathcal{E}_4-\mathcal{E}_2}{\Gamma}
\right]\\
+&
\frac{i\mathcal{E}^2_{\rm H}}{(\mathcal{E}_3-\mathcal{E}_1+2i\Gamma)(\mathcal{E}_3-\mathcal{E}_2+2i\Gamma)}
\left[
\frac{\pi}{2}+
\arctan
\frac{\mathcal{E}_4-\mathcal{E}_3}{\Gamma}
\right].
\end{aligned}
\ee

Due to the energy denominators in Eq.~\eqref{eq_rrr_alldiff}, numerical difficulties
can arise when $\mathcal{E}_1\ne\mathcal{E}_2\ne\mathcal{E}_3\ne\mathcal{E}_1 $ is
not satisfied. Therefore, when $\mathcal{E}_1=\mathcal{E}_2\ne\mathcal{E}_3 $ we
use instead of Eq.~\eqref{eq_rrr_alldiff} the expression
\bege\label{eq_rrr_twosame}
\begin{aligned}
&I_{1}(\mathcal{E}_1,\mathcal{E}_1,\mathcal{E}_3,\mathcal{E}_{4})=\\
&
\frac{\mathcal{E}^2_{\rm H}}{2(\mathcal{E}_1-\mathcal{E}_3)^2}
\log
\left[
\frac{\Gamma^2+(\mathcal{E}_3-\mathcal{E}_4)^2}{\Gamma^2+(\mathcal{E}_1-\mathcal{E}_4)^2}
\right]\\
+&
\frac{\mathcal{E}^2_{\rm H}}{i(\mathcal{E}_1-\mathcal{E}_3)^2}
\left[
\arctan
\frac{\mathcal{E}_4-\mathcal{E}_3}{\Gamma}
-
\arctan
\frac{\mathcal{E}_4-\mathcal{E}_1}{\Gamma}
\right]\\
+&
\frac{\mathcal{E}^2_{\rm H}}{(\mathcal{E}_3-\mathcal{E}_1)(\mathcal{E}_4-\mathcal{E}_1+i\Gamma)}.
\end{aligned}
\ee
Applying $I_{1}(\mathcal{E}_1,\mathcal{E}_1,\mathcal{E}_3,\mathcal{E}_{4})=I_{1}(\mathcal{E}_1,\mathcal{E}_3,\mathcal{E}_1,\mathcal{E}_{4})=I_{1}(\mathcal{E}_3,\mathcal{E}_1,\mathcal{E}_1,\mathcal{E}_{4})$ 
to Eq.~\eqref{eq_rrr_twosame}
one readily obtains expressions for $I_{1}(\mathcal{E}_1,\mathcal{E}_2,\mathcal{E}_3,\mathcal{E}_{4})$
that can be used in the special cases  $\mathcal{E}_1\ne\mathcal{E}_2=\mathcal{E}_3 $ 
or $\mathcal{E}_1=\mathcal{E}_3\ne\mathcal{E}_2 $.

Similarly, when $\mathcal{E}_1=\mathcal{E}_2$,
we do not use Eq.~\eqref{eq_rra_alldiff}, but instead
\bege\label{eq_rra_twosame}
\begin{aligned}
&I_{2}(\mathcal{E}_1,\mathcal{E}_1,\mathcal{E}_3,\mathcal{E}_{4})=\\
&
\frac{\mathcal{E}^2_{\rm H}}{2(\mathcal{E}_1-\mathcal{E}_3-2i\Gamma)^2}
\log
\left[
\frac{\Gamma^2+(\mathcal{E}_3-\mathcal{E}_4)^2}
{\Gamma^2+(\mathcal{E}_1-\mathcal{E}_4)^2}
\right]\\
+&
\frac{i\quad\mathcal{E}^2_{\rm H}}{(\mathcal{E}_1-\mathcal{E}_3-2i\Gamma)^2}
\left[
\frac{\pi}{2}+
\arctan
\frac{\mathcal{E}_4-\mathcal{E}_3}{\Gamma}
\right]\\
+&
\frac{i\quad\mathcal{E}^2_{\rm H}}{(\mathcal{E}_1-\mathcal{E}_3-2i\Gamma)^2}
\left[
\frac{\pi}{2}+
\arctan
\frac{\mathcal{E}_4-\mathcal{E}_1}{\Gamma}
\right]\\
+&
\frac{\mathcal{E}^2_{\rm H}}{(\mathcal{E}_3-\mathcal{E}_1+2i\Gamma)(\mathcal{E}_4-\mathcal{E}_1+i\Gamma)}.
\end{aligned}
\ee

In the special case $\mathcal{E}_1=\mathcal{E}_2=\mathcal{E}_3 $ we use
\bege
\begin{aligned}
I_{1}(\mathcal{E}_1,\mathcal{E}_1,\mathcal{E}_1,\mathcal{E}_{4})=
-\frac{\mathcal{E}^2_{\rm H}}{2(\mathcal{E}_4-\mathcal{E}_1+i\Gamma)^2}.
\end{aligned}
\ee

\acknowledgments
We gratefully acknowledge computing time on the supercomputers
of J\"ulich Supercomputing Center and RWTH Aachen University
as well as financial support from the programme
SPP 1538 Spin Caloric Transport
of the Deutsche Forschungsgemeinschaft.
\bibliography{lasintor}

\end{document}